\documentclass[aps,prb,noshowpacs,twocolumn,showpacs,superscriptaddress,amsmath,amssymb]{revtex4-1}

\usepackage{amsmath}
\usepackage{amssymb}
\usepackage{amsthm}
\usepackage[dvipsnames]{xcolor}

\usepackage{amsbsy}
\usepackage{amstext}
\usepackage{graphicx}
\usepackage{color}
\usepackage{float}
\usepackage{siunitx}
\usepackage{wasysym}

\usepackage{mathtools}
\usepackage{subfigure}
\makeatletter
\usepackage{amsfonts}
\usepackage{dcolumn}
\usepackage{bbold,bm}
\usepackage{soul}
\usepackage{makecell}

\def\ket#1{\mathinner{|{#1}\rangle}}

\newcommand{\vect}[1]{\boldsymbol{#1}}

\makeatother

\begin{document}

\title{Direct driving of electronic and phononic degrees of freedom in a honeycomb bilayer with infrared light}

\author{Martin Rodriguez-Vega}
\email{rodriguezvega@lanl.gov}

\affiliation{Theoretical Division, Los Alamos National Laboratory, Los Alamos, New Mexico 87545, USA}

\author{Michael Vogl}
\email{ssss133@googlemail.com}
\affiliation{Department of Physics, King Fahd University of Petroleum and Minerals, 31261 Dhahran, Saudi Arabia}
\author{Gregory A. Fiete}
\affiliation{Department of Physics, Northeastern University, Boston, MA 02115, USA}
\affiliation{Department of Physics, Massachusetts Institute of Technology, Cambridge, MA 02139, USA}

\begin{abstract}
We study theoretically AB-stacked honeycomb bilayers driven by light in resonance with an infrared phonon within a tight-binding description. We characterize the phonon properties of honeycomb bilayers with group theory and construct an electronic time-dependent tight-binding model for the system following photo-excitation in resonance with an infrared phonon. We adopt an ``atomically adiabatic" approximation, introduced by Mohantya $\&$ Heller PNAS 116, 18316 (2019) to describe classically vibrating nuclei, but obtain the Floquet quasienergy spectrum associated with the time-dependent model exactly. We introduce a general scheme to disentangle the complex low-frequency Floquet spectrum to elucidate the relevant Floquet bands. As a prototypical example, we consider bilayer graphene. We find that light in the low-frequency regime can induce a bandgap in the quasienergy spectrum in the vicinity of the $K$ points even if it is linearly polarized, in contrast with the expectations within the Born–Oppenheimer approximation and the high-frequency regime. Finally, we analyze the diabaticity of the driven electron and driven phonon processes and found contrasting effects on the autocorrelation functions at the same driving frequency: driven phonons preserve the character of the initial state while driven electrons exhibit strong deviations within a few drive cycles.  The procedure outlined here can be applied to other materials to describe the combined effects of low-frequency light on phonons and electrons.
\end{abstract}

\date{\today}
\maketitle

\section{Introduction}

The study of periodically-driven systems has lead to the prediction and discovery of novel phases of matter~\cite{Rudner2020,doi:10.1146/annurev-conmatphys-031218-013423}. For example, Floquet topological insulators~\cite{oka2009,lindner2011}, discrete time crystals \cite{Yao_2017,choi2017observation}, hidden order phases\cite{Stojchevska177}, Kapitza pendulum-like many-body phases \cite{PhysRevB.100.104306}, novel topological phases\cite{Potter_2016,PhysRevLett.116.250401,PhysRevB.100.085138} such as higher order Floquet topological phases~\cite{PhysRevLett.123.016806,Huang_2020,PhysRevB.100.085138,PhysRevB.99.045441,PhysRevResearch.2.013124,PhysRevB.102.094305}, and emergent Weyl semimetals and Fermi arcs~\cite{PhysRevB.96.041126,Zhu_2020,Hubener2017,PhysRevLett.121.036401,PhysRevB.94.235137,PhysRevB.99.115136}. Optical cavities provide another pathway to realize light-driven phases of matter ~\cite{Hubener2020,PhysRevB.99.235156,MARTIN2019101}. Experimentally, Floquet states have been observed in driven topological insulators via time-resolved photo-emission spectroscopy~\cite{Wang453,Mahmood2016}, and the light-induced anomalous Hall effect has been reported in graphene~\cite{mciver2018lightinduced,sato2019}. Floquet states have also been reported in photonic systems~\cite{Rechtsman2013,PhysRevLett.122.143903,Mukherjee856,PhysRevX.5.011012,Maczewsky2017,Mukherjee2017}.

Many of the theoretical works have focused on the high-frequency regime, where analytical tools are available to derive effective models \cite{Abanin_2017,Eckardt_2015,PhysRevB.93.144307,PhysRevA.68.013820,bukov2015universal,BLANES2009151,Magnus1954}. However, the low-frequency regime is potentially more relevant for experimental applications since it allows driving without electron resonances with high-lying states. Recent theoretical developments ~\cite{vogl2020effective,PhysRevX.9.021037,Rodriguez_Vega_2018} allow us to obtain non-perturbative effective Hamiltonians and construct Floquet states perturbatively. Studies in the low-frequency regime have revealed breaking of Thouless pumping~\cite{Privitera2018}, emergence of Weyl semimetal states~\cite{PhysRevB.96.041126,PhysRevB.99.115136}, and a plethora of large topological invariants~\cite{PhysRevB.90.195419,PhysRevLett.121.036402,PhysRevLett.110.200403,PhysRevLett.111.136402,PhysRevB.93.144307,Li_2018, PhysRevB.89.205408,PhysRevB.87.201109,PhysRevLett.111.047002,PhysRevB.82.235114,PhysRevX.3.031005}.

When considering low-frequency light driving an electronic system, the light can be in resonance with a phonon, which for materials typically possesses excitation frequencies in the THz regime. However, the effect of the light on the phonons and electrons is usually not treated in a unified fashion. Recently, the phonons have been considered in a Floquet picture revealing side-band structure in phonon-dressed states in graphene~\cite{Hubener2018}, phonon-dressed spins carrying a net out-of-plane magnetization in MoS$_2$~\cite{Shin2018}, and phonon-induced Floquet second-order topological phases~\cite{chaudhary2019phononinduced}.

In this work, we study the effect of low-frequency light in honeycomb bilayers taking into account both the electronic and phononic degrees of freedom within a time-dependent and periodic tight-binding model. We characterize the lattice vibrations using a group theory approach, and construct the time-dependent model based on the symmetry-allowed real-space lattice displacements corresponding to an infrared phonon's irreducible representation. The direct interaction of the electrons with the laser pulse is introduced via minimal coupling.

In the derivation of our time-dependent tight-binding model, we adopt the ``atomically adiabatic" approximation introduced in Ref.~[\onlinecite{Mohanty18316}] to describe lattice vibrations in graphene associated with thermal excitations. In this approximation, the in-plane atomic orbitals follow the nuclei adiabatically, but the $p_z$ orbitals are treated exactly within the tight-binding model. The atomically adiabatic approximation is distinct from the adiabatic Born-Oppenheimer approximation, which has been shown to fail to describe electron dynamics in graphene~\cite{Pisana2007,Mohanty18316}. We note that since we do not consider the back-action of the dynamical electronic states on the phonons, inelastic scattering constitutes an important correction to the work here presented.

As a prototypical example, we consider AB-stacked bilayer graphene driven with infrared light in resonance with a phonon. We show that the low-frequency Floquet quasienergy spectrum develops a gap in contrast with the expectation from an adiabatic  Born-Oppenheimer approximation. The procedure outlined here can be applied to other van der Waals materials to describe the combined effects of low-frequency light on phonons and electrons.

The remainder of the paper is organized as follows. In section \ref{secGT} we perform a group theory analysis of the lattice vibrations, which apply to honey bilayers with $D_{3d}$ point group. We discuss the symmetry properties of the lattice that allow coherent photo-excitation. Based on the phonon properties, we construct a tight-binding model that captures the lattice distortions corresponding to an infrared $E_u$ phonon mode in section \ref{sec:tightb}. In section \ref{sec:pd}, we describe the laser excitation of the phonons and  derive the effective time-dependence in the long-time limit. In section \ref{sec:graphene}, we apply our procedure to bilayer graphene, perform first principles calculations to determine the energy scales not accessible from the symmetry arguments of group theory, and solve the corresponding Floquet problem exactly. Finally, in section \ref{sec:conclusions}, we present our conclusions.

\section{Group theory analysis of the lattice vibrations}
\label{secGT}

\begin{figure}[t]
	\begin{center}
		\includegraphics[width=7.0cm]{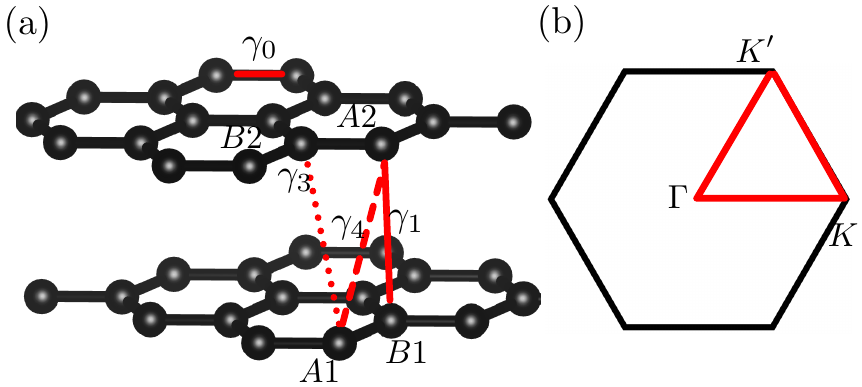}
		\caption{(Color online) 
		(a) Bilayer graphene lattice structure. The red lines indicate the hopping between lattice sites. 
		(b) Brillouin zone with a highlighted high symmetry path.}
		\label{fig:bands_gga}
	\end{center}
\end{figure}

We start studying the group theory aspects of the lattice vibrations. We assume that the honeycomb bilayer is AB-stacked and has $D_{3d}$ point group. In Fig. \ref{fig:bands_gga}, we show the lattice labeling the relevant hopping amplitudes and sublattices, and the Brillouin zone (BZ). The matrix representation for the twelve symmetry operations of the space group are presented in Appendix \ref{app:br}. Since there are $N=4$ atoms in the primitive unit cell located at Wyckoff positions $2c$ ($(0,0,z), (0,0,-z)$) and $2d$ ($(1/3,2/3,z), (2/3,1/3,-z)$), we have $3 N = 12$ phonon modes at the $\Gamma$ point. The lattice vibration representation is given by $\Gamma_{\rm{latt.vib.}}= 2 {A}_{1 {g}} \oplus 2 {A}_{2 {u}}\oplus2 {E}_{{g}}\oplus2 {E}_{{u}}$, which we obtain using the Bilbao Crystallographic Sever (BCS)~\cite{kroumova2003}. Six modes are even under inversion (optical Raman active $A_{1g}$ and $E_g$ modes), and six modes are odd under inversion ($A_{2u}, E_u$), including three acoustic modes and three optical modes. 

In order to determine the set of lattice vibrations that block-diagonalize the dynamical matrix, we construct the projection operators~\cite{GroupTheoryDress2008, gtpack1, gtpack2} 
$
\hat P^{(\Gamma_n)}_{kl} = \frac{l_n}{h} \sum_{C_\alpha} \left( D_{kl}^{(\Gamma_n)}(C_\alpha) \right)^* \hat P(C_\alpha)$,
where $\Gamma_n$ are the irreducible representations, $C_\alpha$ are the elements of the group, $D_{kl}^{(\Gamma_n)}(C_\alpha) $ is the irreducible matrix representation of element $C_\alpha$, $h$ is the order of the group, and $l_n$ is the dimension of the irreducible representation. The resulting vectors are shown in Table \ref{tab:GT-modes}, derived with \small{ISODISTORT}~\cite{hatch2003s}. We obtain that shear modes, where the layers move uniformly in opposite in-plane directions are allowed by symmetry and possess irreducible representation (irrep) $E_g$. Breathing modes with irrep $A_{1g}$, where the layers move away and towards each other are also allowed. This mode can be constructed by subtracting the carbon atoms at Wyckoff position 2c from the 2d carbon atoms.  Next, we will discuss the electronic tight-binding Hamiltonian. 

\begin{table}[]
\begin{tabular}{|c|c|c|c|c|}
\hline          
& \multicolumn{2}{c|}{2d} & \multicolumn{2}{c|}{2c} \\ \cline{2-5} 
                               & $A_1$  & $B_2$  & $B_1$  & $A_2$  \\ \hline
\multicolumn{1}{|c|}{$A_{1g}$} & $(0,0,1)$  & $(0,0,-1)$ & $(0,0,-1)$ & $(0,0,1)$  \\ \hline
$A_{2u}$                       & $(0,0,1)$  & $(0,0,1)$  & $(0,0,1)$  & $(0,0,1)$  \\ \hline
$E_g$                          & $(-1,0,0)$ & $(1,0,0)$  & $(1,0,0)$  & $(-1,0,0)$ \\ \hline
$E_u$                          & $(-1,0,0)$ & $(-1,0,0)$ & $(-1,0,0)$ & $(-1,0,0)$ \\ \hline
\end{tabular}
\caption{Displacements that block-diagonalize the dynamical matrix. The top row labels the Wyckoff position. The second row indicates the carbon atoms sublattice label. The vectors indicate the direction of the displacements in Cartesian coordinates, with the $x-$axis along the $a$ crystallographic axis. For the doubly degenerate $E$ modes, the other partner can be constructed by orthogonality.}
\label{tab:GT-modes}
\end{table}

\section{Static tight-binding model}
\label{sec:tightb}

The spinless tight binding Hamiltonian for an undistorted AB honeycomb bilayer can be written as~\cite{McCann_2013}
\begin{equation}
\begin{aligned}
\mathcal H=&-\sum_{\vect R}\sum_{n=1}^3 \gamma_0(\vect \delta_{n,1}^2) a^\dag_{1}(\vect R)b_1(\vect R+\vect \delta_{n,1})\\
&-\sum_{\vect R}\sum_{n=1}^3 \gamma_0(\vect \delta_{n,2}^2) b^\dag_{2}(\vect R)a_2(\vect R+\vect \delta_{n,2})\\
&-\gamma_3 \sum_{\vect R} a_1^\dag (\vect R)b_{2}(\vect R+\vect \delta_{n,1})\\
&-\gamma_1 \sum_{\vect R} a_2^\dag (\vect R)b_{1}(\vect R)+h.c.,
\end{aligned}
\label{TB_Ham}
\end{equation}
where only intra-layer nearest neighbor hopping with amplitude $\gamma_0$ is considered. Here, $\gamma_1$ describes the interlayer tunneling amplitude from the atoms lying directly on top of each other: $A2 \leftrightarrow B1$, and $\gamma_3$ is the tunneling amplitude from sublattices $B2 \leftrightarrow A_1$. In principle, $\gamma_3$ should be $\delta_{n,1}^2$-dependent to treat it on the same level as $\gamma_0$. However, $\gamma_3$ is already small compared with $\gamma_0$, and the distortions are assumed to be small, which leads to higher order corrections that we neglect. 

In Eq.\eqref{TB_Ham}, $a_l$  ($b_l$) is the annihilation operator for electrons on sublattice $A (B)$ and layer $l=1,2$, and $\vect R$ labels the real space positions of the lattice sites on sublattice $A$ of each layer. The nearest-neighbor vectors at site $\vect R$ for layer $l$ can be written as 
\begin{equation}
	\vect \delta_{n,l}=(-1)^la \mathcal R\left(\frac{2\pi n}{3} \right) \hat y,
\end{equation}
which are labeled by the index $n=1,2,3$, $\mathcal R(\theta)$ is a rotation matrix with angle $\theta$ and $a$ is the nearest neighbor lattice spacing. To be flexible enough to accurately describe distortions to the lattice we  consider a distance $\delta$ dependent hopping function modeled as \cite{Rost_2019}
\begin{equation}
	\gamma_0(\vect \delta^2)=c_1e^{-c_2\vect \delta^2},
\end{equation}
where the parameters $c_i$ can be found via a fit to combined set of first, second, and third nearest hopping parameters\cite{PhysRevB.66.035412,kundu}.

Now we analyze the changes for the case of the lattice distortion induced by driving one of the in-plane IR phonon modes with irrep $E_u$, constructed as discussed in Sec. \ref{secGT}. We may choose a coordinate system where only the nearest neighbor vectors are affected, provided the shift is sufficiently small. The shift then is in the direction $\vect \Delta$. Consequently, the nearest neighbor vectors become 
\begin{equation}
		\vect \delta_{n,l}(\mathcal A)=(-1)^la \mathcal R\left(\frac{2\pi n}{3} \right)\hat y+ \frac{\Delta}{2} \hat \Delta.
\end{equation}
where $\Delta$ is the shift amplitude and the factor $1/2$ was introduced because the shift in the coordinate system, where only the nearest neighbor vectors change, is twice of the shift of the $E_u$ mode.  Now we replace the new nearest-neighbor vectors into the Hamiltonian Eq.\eqref{TB_Ham} to find the single particle Hamiltonian
\begin{equation}
	\mathcal H=\begin{pmatrix}
	0&f_{1}(\vect k, \vect \Delta)&0&\gamma_3 g^*(\vect k, \vect \Delta) \\
	f_{1}^*(\vect k,\vect \Delta)&0&\gamma_1&0\\
	0&\gamma_1&0&f_{2}(\vect k,\vect \Delta)\\
	\gamma_3 g(\vect k,\vect \Delta)&0&f^*_{2}(\vect k, \vect \Delta)&0
	\end{pmatrix},
\end{equation}
where the geometric factor for layer $l=1,2$ is 
\begin{equation}
	f_l( \vect \Delta)=\sum_{n=1}^3 \gamma_0(\vect \delta_{n,l}^2)e^{i\vect k \cdot \vect \delta_{n,l}(\vect \Delta)},
\end{equation}
and $g(\vect k,\vect \Delta)=\sum_{n=1}^3 e^{i\vect k \cdot \vect \delta_{n,1}(\vect \Delta)}$.  It is important to recognize that introducing time-dependent lattice sites means that the tight binding model now implicitly works with a time dependent Wannier basis. This fact has important consequences.

We recall that to construct a tight binding model one usually works as follows.  One starts from the full Hamiltonian $H$ and then chooses a set of Wannier orbitals to project on. \cite{PhysRev.133.A171}  This procedure can be described by a projection operator $P$. For the time-independent case this makes it possible to compute an effective tight binding Hamiltonian $H_{\mathrm{TB}}=PHP$. 

For the time-dependent case we recall that, in principle, one has to project the full Schr\"odinger equation including the time derivative, $-i\partial_t$. For the time independent case this causes no problem because $[\partial_t,P]=0$. For the time-dependent case, however, one has to keep track of this term and one finds a correction term $H_{\mathrm{TB}}(t)=i(\partial_t P)P+PHP$, which is similar to the result in Ref.~[\onlinecite{PhysRevLett.110.065301}]. 

One should recognize that for a Floquet system this term is relevant in the high frequency regime. Neglecting it in the high frequency regime can lead to nonphysical results, such as broken translation symmetry in momentum space. A demonstration of this effect is given in appendix \ref{app:dimer}, where we consider a simple dimer model which shows how this effect unravels analytically. The same effect appears in the high-frequency regime for our current model. 

\begin{figure}[h]
	\begin{center}
		\includegraphics[width=6.5cm]{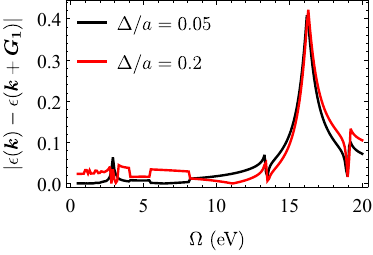}
		\caption{(Color online) Quasienergy difference $|\epsilon(\boldsymbol k  )-\epsilon(\boldsymbol k + \boldsymbol G_1 )|$ as a function of drive frequency for two lattice distortion amplitudes $\Delta/a$. For small-enough distortions, transnational symmetry in momentum space is recovered in the low-frequency regime.}
		\label{fig:val}
	\end{center}
\end{figure}

In the general context, in order to obtain a physical approximation, one must restrict to frequencies satisfying the condition $\hbar \Omega \ll W $, where $W$ is the bandwidth of the system. In particular, here we propose to diagnose the regime of validity by comparing the quasienergies at a high-symmetry point in the BZ ($\epsilon(\boldsymbol k)$) with the quasienergy at a momentum shifted by a reciprocal lattice vector ($\epsilon(\boldsymbol k + \boldsymbol G_1)$). This criterion for validity is motivated by the observation that if this difference is negligible, we have recovered translational symmetry in momentum space. Translational symmetry is broken for high frequency regimes where our approximation is not valid. As Fig. \ref{fig:val} shows for a representative set of parameters, as the drive frequency decreases, the difference approaches zero for small enough lattice distortions characterized by $\Delta/a$. We note that our estimates, as explained later in the text, suggest that experimentally one can achieve $\Delta/a < 0.05$ and the phonon frequency is $\Omega \approx 0.2$~eV. Therefore, the conditions of the phonon of interest in this work are well approximated neglecting the time dependence of the projection operator. This diagnostic procedure could be implemented in other non-interacting Floquet system with driven phonons to establish the theory validity regime.

\section{Effect of the infrared laser drive and time-dependent Hamiltonian}
\label{sec:pd}

When the laser is incident on the system normal to the surface, the time-dependent Hamiltonian can be written as
\begin{widetext}
\begin{equation}
	\mathcal H(t) =\begin{pmatrix}
	0&f_{1}(\vect k(t),\vect \Delta(t))&0&\gamma_3 g^*(\vect k(t), \vect \Delta(t)) \\
	f_{1}^*(\vect k(t),\vect \Delta(t))&0&\gamma_1&0\\
	0&\gamma_1&0&f_{2}(\vect k(t),\vect \Delta(t))\\
	\gamma_3 g(\vect k(t),\vect \Delta(t))&0&f^*_{2}(\vect k(t), \vect \Delta(t))&0
	\end{pmatrix},
	\label{eq:ham1_result}
\end{equation}
\end{widetext}
where $\vect k (t)  = \vect k - \vect A(t)$, and $ \vect A(t)$  is the vector potential. This form arises from coupling of the light with the electronics degrees of freedom, valid for not too strong drives \cite{PhysRevB.101.205140}. The time-dependent model Hamiltonian Eq.\eqref{eq:ham1_result} for both driven electrons and phonons is the first result of this work. 

The time-dependence of the lattice distortion $\Delta(t)$ is derived through the the potential functional governing the IR phonon $Q_{\text{IR}}$ dynamics after photo-excitation and is given by \cite{baroni2001,forst2011} 
\begin{align}
&V[Q_{\text{IR}}] =  \frac{1}{2}\Omega^2_{\text{IR}} Q_{\text{IR}}^2+\vect{Z}^* \cdot \vect{E}_0 \sin(\Omega t) F(t) Q_{\text{IR}}.
\label{eq:non-linear-pot_main}
\end{align}
where $\vect{E}_0$ is the electric field amplitude and $Z^*$ is the mode effective charge~\cite{gonze1997, baroni2001,Bistoni_2019}, which determines the coupling strength of the IR modes with the laser~\cite{buhrer1986phonons,Giannozzi1994}. 

In the most general case, the effective charge depends on frequency\cite{Bistoni_2019,trickey1990density},
$
Z(\omega)=Z^{st} +Z^{dyn}(\omega).
$ 
The first term, $Z^{st}=V/(|e|) \partial \vec{P}/\partial \vec{u}^I$, corresponds to the static contribution where $V$ is the area of the unit cell, $e$ is the electron charge, $\vec P$ is the dipole moment, and $\vec u^I$ is the ion displacement. However, $Z^{st}$ is well-defined only for insulators.  The second dynamic contribution, $Z^{dyn}(\omega)$, becomes relevant when the electronic gap is smaller than the phonon frequencies such as in bilayer graphene~\cite{Bistoni_2019,binci2020firstprinciples}. It originates from the polarization of the valence electrons induced by the atomic displacements. $Z^{dyn}(\omega)$ can be calculated within advanced DFPT schemes~\cite{RevModPhys.73.515,Bistoni_2019, trickey1990density}. Alternatively, we could adopt experimental values. 

In the following we assume that the laser profile has the Gaussian form $F(t)=\exp\{ -t^2/(2 \tau^2)\}$, where $\tau^2$ is the variance. The differential equation governing the dynamics of $Q_{\text{R}(i)}$ becomes $\partial^2_t Q_{\text{R}(i)} + \eta \partial_t Q_{\text{R}(i)}   = -\partial_{Q_{\text{R}(i)}} V[Q_{\text{IR}},Q_{\text{R}(i)}]$, where $\eta$ is a damping constant. In order to obtain the amplitude of the displacement for a given applied laser, we solve the equation of motion numerically. The normal mode $Q_{\text{R}}(t)$ is related to the real-space displacement of atoms in the $j$-direction via $\Delta_{j}(t) = Q_{\text{IR}}(t)q_{j}/\sqrt{m_C} $, where $q_{j}$ is the normalized dynamical matrix eigenvector for the carbon atoms obtained from first-principles calculations.

In the absence of damping, and for laser frequency in resonance with the $E_u$ modes, $\Omega = \Omega_{\text{IR}} $, and $\Omega_{\text{IR}} \tau \ll 1$, the solution is given by 
\begin{eqnarray}
Q_{\text{IR}}(t) &=& \sqrt{2 \pi} Z^* E_0 \tau/\Omega_{\text{IR}} \cos \left(\Omega_{\text{IR}} t \right) \sinh\left[ (\Omega_{\text{IR}} \tau)^2 \right]e^{-(\Omega_{\text{IR}} \tau)^2} \nonumber\\ &\equiv& Q_{max} \cos \left(\Omega_{\text{IR}} t \right),
\end{eqnarray}
where we assume~\cite{subedi2014} $Q_{\text{IR}}(-\infty)=\partial_t Q_{\text{IR}}(-\infty)=0$, and $Q_{max} = \sqrt{2 \pi} Z^* E_0 \tau/\Omega_{\text{IR}} \sinh\left[ (\Omega_{\text{IR}} \tau)^2 \right]e^{-(\Omega_{\text{IR}} \tau)^2}$ as boundary conditions. Therefore, the maximum amplitude of the phonon scales linearly with the applied electric field. 
 
For simplicity, we assume the long-time time-dependence $\vect \Delta(t) = \vect \Delta \sin \left(\Omega_{\text{IR}} t \right)$. Furthermore, the sinusoidal time-dependence ensures that the undriven system is recovered for no only $\Omega \rightarrow 0$ but also $t \rightarrow 0$. For the electronic drive, we consider the vector potential $\vect A(t) = \tilde A \hat \Delta \sin\left(\Omega_{\text{IR}} t \right)$, where $\tilde A \equiv (e/\hbar) a A = e  E/(\hbar \Omega_{\text{IR}} ) $ is the dimensionless parameter that defines the coupling strength with the electrons. The factor $\hat \Delta$ arises from aligning the polarization with the phonon displacement direction. 

For periodically driven systems,  $\mathcal H(t) = \mathcal H(t+2\pi/\Omega )$. such as the one defined here, and we can employ Floquet theory to study the system. The remaining discrete time-translation symmetry allows one to use the Floquet theorem~\cite{floquet}  to write the wave functions as $|\psi(t)\rangle= e^{i \epsilon t}|\phi(t)\rangle,$ where $|\phi(t+2 \pi / \Omega)\rangle=|\phi(t)\rangle$ and $\epsilon$ is the quasienergy. This wavefunction obeys the Floquet-Schr\"odinger equation 
\begin{equation}
\left[H(t)-i \partial_{t}\right]|\phi(t)\rangle=\epsilon|\phi(t)\rangle.
\end{equation}
The exact solution can be written formally as $U_{F}=\mathrm{T} \exp \left\{-i \int_{0}^{2 \pi / \Omega} H(s) d s\right\}=e^{-i H_{F} T}$. We employ the extended-state picture which relies on an expansion of the steady states in a Fourier series  $|\phi(t)\rangle=\sum_{n} e^{i n \Omega t}\left|\phi_{n}\right\rangle$  which leads to $\sum_{m}\left(H^{(n-m)}+\delta_{n, m} \Omega m\right)\left|\phi_{m}\right\rangle=\epsilon\left|\phi_{n}\right\rangle$. The Hamiltonian Fourier modes are given by $H^{(n)}=\int_{0}^{2 \pi} d \tau /(2 \pi) H(\tau) e^{-i \tau n}$. In the next section, we discuss the application to bilayer graphene and discuss the effects in the electronic band structure.

\section{Application to bilayer graphene}
\label{sec:graphene}

Now, we apply the procedure discussed before to bilayer graphene, which  belongs to the space group $P\bar 3 m 1$ (No. 164) with $D_{3d}$ point group at the $\Gamma$ point~\cite{PhysRevB.79.125426}. Experimental evidence indicates that the Frozen phonon picture (adiabatic limit) fails to describe the phonon properties in graphene as a function of the carrier density because the electron relaxation time is larger than the phonon frequency.~\cite{Pisana2007} However, in bilayer graphene, time- and angle-resolved photoemission spectroscopy experiments  revealed carrier dynamics that were explained within the frozen phonon picture.~\cite{PhysRevLett.114.125503}. Also, enhancement of the electron-phonon coupling in bilayer graphene after phonon excitation has been reported~\cite{PhysRevB.95.024304}. Theoretical studies in graphene~\cite{Hubener2018} and transition metal dichalcogenides~\cite{Shin2018} employing first principles calculations indicate that the electronic structure of phonon-driven systems can be captured with Floquet theory. In this section, we analyze the electronic dynamics within our time-dependent tight-binding model.

\begin{figure}[t]
	\begin{center}
		\includegraphics[width=8.5cm]{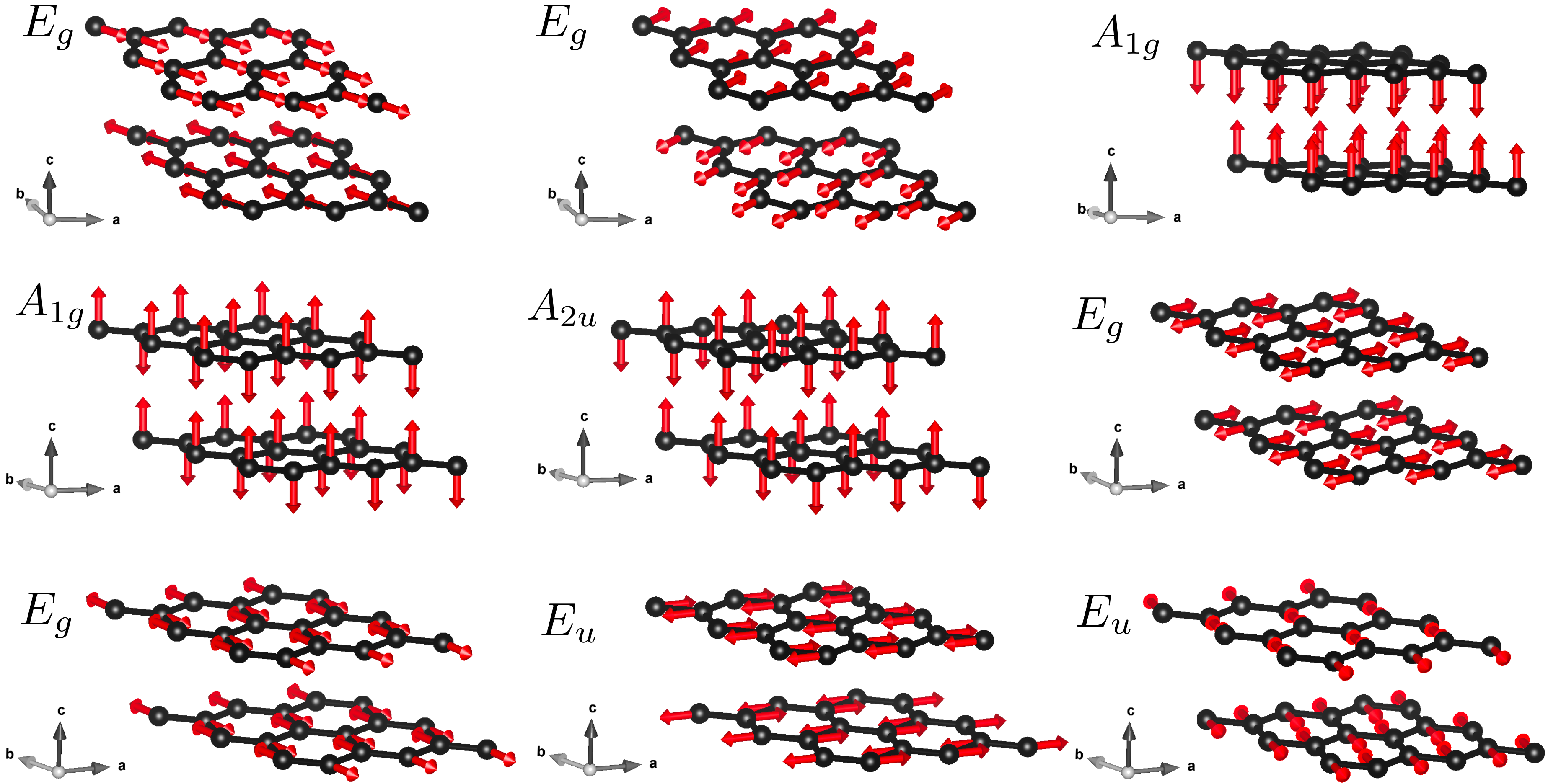}
		\caption{(Color online) Bilayer graphene $\Gamma$-point real-space lattice vibrations. The low-frequency shear modes (top-left corner) are relevant for the non-linear phonon processes.  These figures were created with VESTA.}
		\label{fig:real_space}
	\end{center}
\end{figure}

First, we calculate the phonons frequencies and eigenmodes at the $\Gamma$ point. It is enough to consider regions near the $\Gamma$ point because photons carry momentum that is small compared to the size of the Brillouin zone. For this, we employ density-functional perturbation theory (DFPT)~\cite{RevModPhys.73.515}, imposing the acoustic sum rule to the dynamical matrix, as implemented in {\sc QUANTUM ESPRESSO}~\cite{QE-2017, QE-2009, doi:10.1063/5.0005082}. Table \ref{tab:phonons} shows the phonon frequencies, while the lattice displacements are shown in Fig. \ref{fig:real_space}. The first observation is that the DFPT lattice displacements are consistent with the group theory results. We also compute the electronic band structure within density functional theory (DFT). The details of the calculation are presented in Appendix \ref{sec:bands} . In Fig. \ref{fig:irreps}, we show the bands along a high-symmetry path with their corresponding irreducible representations at the special high-symmetry points $\Gamma$, K, and M obtained with \small{IrRep}~\cite{iraola2020irrep}.  We note that adiabatic and nonadiabatic phonon frequencies can be computed following the approach introduced in Ref. [\onlinecite{PhysRevB.82.165111}].

\begin{figure}[t]
	\begin{center}
		\includegraphics[width=8.5cm]{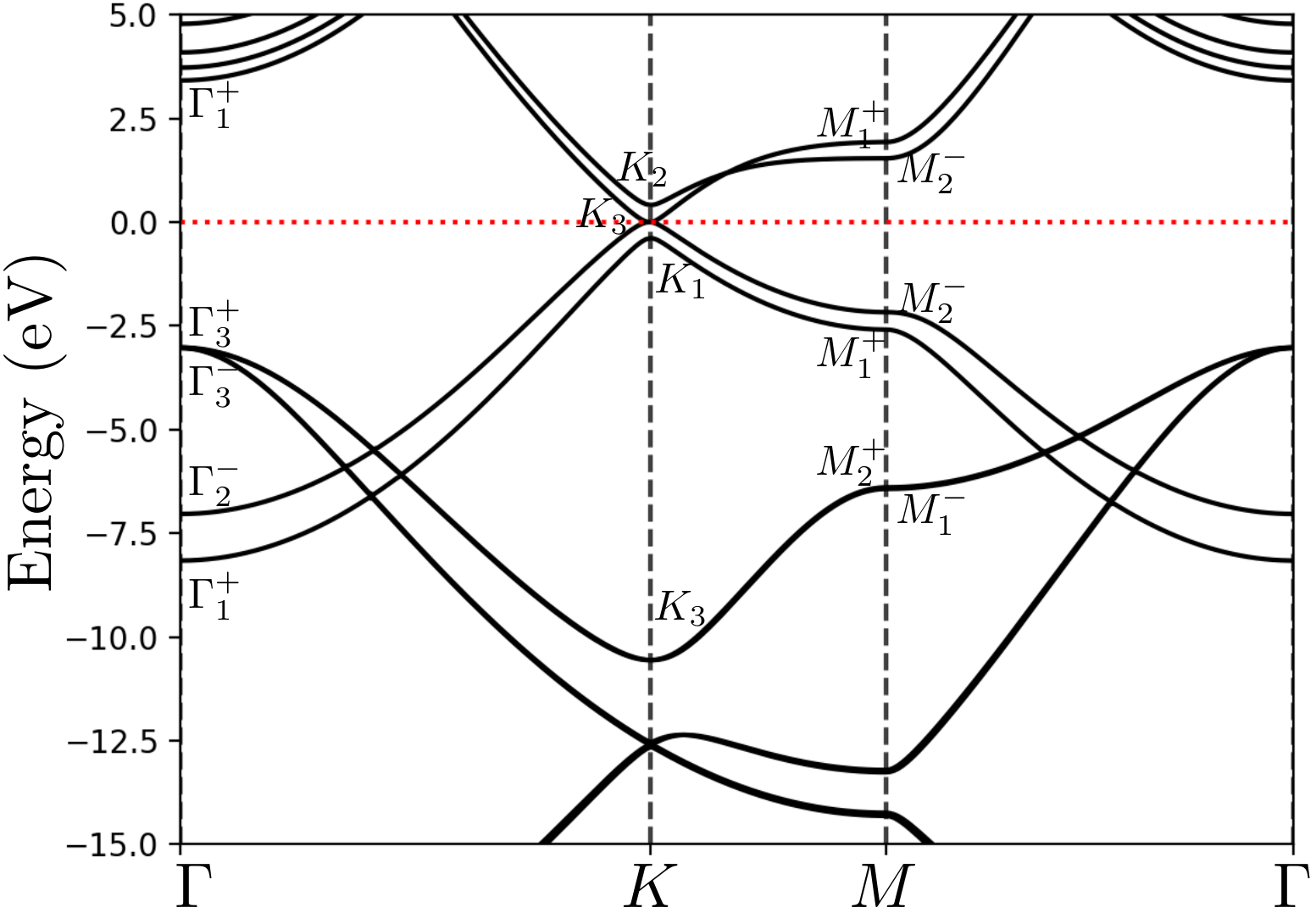}
		\caption{(Color online) AB-stacked bilayer graphene band structure with corresponding irreducible representations at the high-symmetry points $\Gamma$, $K$, and M. The symmetry analysis was performed with \small{IrRep}~\cite{iraola2020irrep}. }
		\label{fig:irreps}
	\end{center}
\end{figure}

\begin{table}[b]
\begin{tabular}{|c|c|c|}
\hline
      & Frequency (THz) & irrep    \\ \hline
1     & 0               & $A_{2u}$ \\ \hline
2,3   & 0               & $E_u$    \\ \hline
4,5   & 0.581536        & $E_g$    \\ \hline
6     & 2.854691        & $A_{1g}$ \\ \hline
7     & 26.045346       & $A_{1g}$ \\ \hline
8     & 26.122256       & $A_{2u}$ \\ \hline
9,10  & 46.873384       & $E_g$    \\ \hline
11,12 & 47.027640       & $E_u$    \\ \hline
\end{tabular}
\caption{$\Gamma$ point phonon modes for bilayer graphene without SOC. }
\label{tab:phonons}
\end{table}

\begin{figure}[t]
	\begin{center}
		\includegraphics[width=6.5cm]{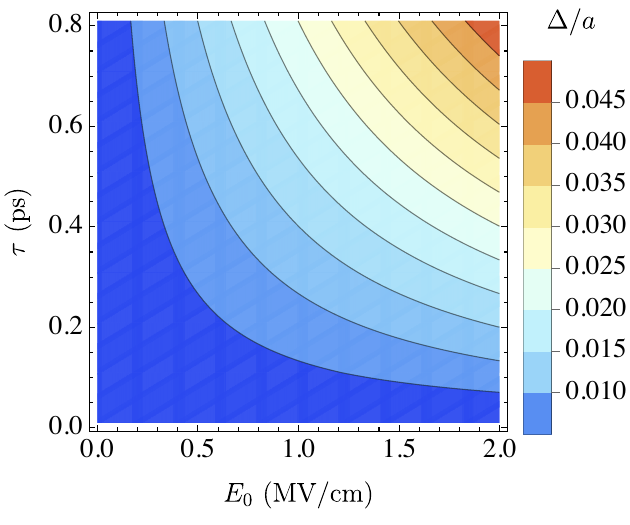}
		\caption{(Color online) $\Delta/a$ as a function of the laser pulse width $\tau$ and the peak electric field $E_0$, where $a \approx 0.142$ nm is the interlayer carbon-carbon distance.}
		\label{fig:color_max_ampl}
	\end{center}
\end{figure}

We will drive the $E_u$ mode with frequency $\Omega= 0.2$~eV, with  $\hat \Delta = \vect \Delta/|\vect \Delta| \approx (0.049, -0.998)$. The next step is to calculate the Born effective charges, which determine the coupling strength between the phonon and the incident laser. Here, we will adopt the value measured by Kuzmenko \textit{et. al.} in Ref.~\onlinecite{Kuzmenko2009} for the $E_u$ mode with frequency  $\Omega= 0.2$~eV, $Z^*~\approx 0.25 e/\sqrt{m_C}$, where $e$ is the electron charge and $m_C$ is the carbon mass, similar to the equivalent mode in graphite~\cite{NEMANICH1977117}. With the  Born effective charges for bilayer graphene, we can compute the relative displacement induced for a given laser peak electric field $E_0$ and laser pulse width $\tau$. Away from the regime $\Omega_{\text{IR}} \tau \ll 1$, we need to solve the equation of motion derived from the potential \eqref{eq:non-linear-pot_main} numerically. In Fig. \ref{fig:color_max_ampl}, we plot $\Delta/a$ as a function of the peak electric field $E_0$ for several values of the laser pulse width $\tau$. The maximum amplitude still scales linearly with the applied electric field $E_0$ in this regime. In particular, for $\tau=0.8$~ps and peak electric field $\sim 1 $MV$/$cm we can reach distortions of about 5$\%$ of the carbon-carbon distance ($\Delta/a \approx 0.05$). We assume that the effects of non-linear phonon couplings~\cite{forst2011,mitrano2016,FORST201324,subedi2014,juraschek2017,Juraschek2017b,subedi2017,Juraschek2018,Juraschek2019,Juraschek2020,juraschek2019phonomagnetic,kalsha2018,gu2018,rodriguezvega2020phononmediated, PhysRevX.10.031028,PhysRevLett.125.137001} are negligible.

In the tight-binding model,  we use the nearest neighbor, next nearest neighbor and third nearest neighbor distances $a,\sqrt{3}a,2a$ and we find the coefficients $c_1\approx12$ eV and $c_2\approx1.5 a^{-2}$. For the interlayer tunneling, we use $\gamma_1= 0.361$~eV and $\gamma_3 = 0.283$~eV.~\cite{jung2014} These parameters fully define the time-dependent Hamiltonian. In the next sections, we discuss a low-energy approximation for the tight-binding model, and solve the full model exactly within a Floquet scheme in the extended space.

\subsection{Low-energy approximation}

Before we proceed with the solution in the low-frequency regime, and to get intuition about the model, we consider the Hamiltonian near the $K$ point and for small lattice displacements $\vect \Delta$. We find that the effective Hamiltonian is given by

\begin{equation}
    H(t)\approx \begin{pmatrix}
    \hbar v_F(\vect k+\vect B(t)+\vect A(t))\vect \sigma&T(\vect k+\vect A(t))\\
    T^\dag(\vect k+\vect A(t))&\hbar v_F(\vect k-\vect B(t)+\vect A(t))\vect \sigma
    \end{pmatrix},
\end{equation}

where $\hbar v_F=3/2a\gamma_0=3/2ac_1e^{-a^2c_2}$. The diagonal blocks describe the graphene layers with linearized bands near the $K$ point. They include the usual electromagnetic vector potential $\vect A(t)$ introduced via minimal substitution as well as an additional vector potential $\vect B(t)=c_2(\Delta_2,\Delta_1)$ (note $(\Delta_1,\Delta_2)=\Delta \hat\Delta$)  induced by lattice distortion. The additional vector potential $\vect B(t)$ has different signs for different layers. Lastly, the off-diagonal blocks
\begin{equation}
    T(\vect k)=\begin{pmatrix}0&\hbar v_3(k_x+ik_y)\\
    \gamma_1&0\end{pmatrix},
\end{equation}
where $\hbar v_3=3/2a\gamma_3$, describe the  interlayer hopping including trigonal warping. Now, we proceed with the solution of the low-frequency Floquet problem. 

\subsection{Exact quasienergy spectrum}

To study the effect of infrared light in bilayer graphene, we solve the Floquet-Schr\"odinger equation in extended space exactly. Within this approach, the low frequency regime involves many overlapping Floquet modes, which leads to a dense quasienergy spectrum with no easy comparison with the original band structure. This, however, can be avoided if one chooses to plot bands for judiciously chosen Floquet copies. After all, only one Floquet copy is necessary to describe the physical problem.  Treating the driven problem for $N$ Floquet copies, we recall that one finds an eigenvector of the form $\psi=(\psi_{-N/2},...,\psi_0,\psi_{N/2})$, where $\psi_m$ is the wavefunction corresponding to the sector of Floquet copy $m$. 

For the undriven system every solution $\psi$ will have only one of the $\psi_m\neq 0$. In this case, the spectrum for Floquet copy zero for a Hamiltonian of dimension $d$ can be obtained if we pick all $d$ eigenvectors with spectral weight $|\langle\psi_0|\psi_0\rangle|^2=1$. For the driven case, if the drive is sufficiently weak, the spectral weight $|\langle\psi_0,\psi_0\rangle|^2$ will be especially large for $d$ of the eigenvectors. In this case one may plot the quasi-energies corresponding to these eigenvectors and may obtain a band structure that can easily be compared to the bandstructure of the undriven system. This is similar to the approach taken in Ref.[\onlinecite{Katz_2020}], where instead of taking the $d$ largest contributions, contributions up to an arbitrary cut-off in the averaged density of states were taken. Our approach is more robust and can be used to check for convergence of the Floquet band structure. All our results are converged with respect to the number of Floquet modes $N$, and typically we require $N \gtrapprox 20$. 

\begin{figure}[t]
	\begin{center}
		\includegraphics[width=8.5cm]{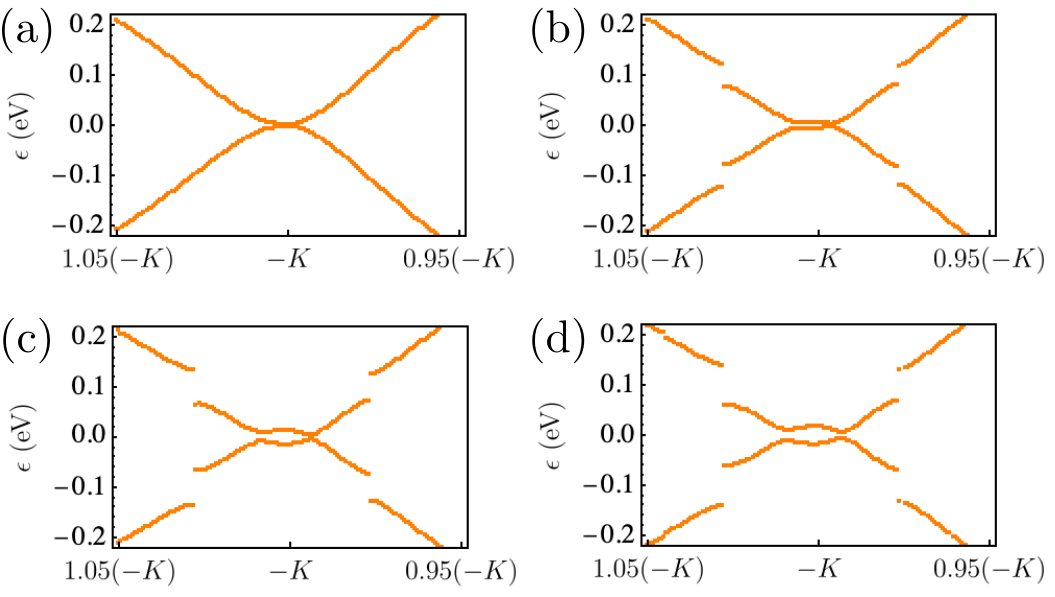}
		\caption{(Color online) Light-driven bilayer graphene exact quasienergies in the vicinity of the $-K$ point. The pulse duration is $\tau = 0.3$~ps and the frequency $\Omega = 0.2$~ eV. The peak electric field intensities $E_0$ are (a) $0.0$, (b) $0.2$,(c) $0.3$, and (d) $0.35$ MV/cm.}
		\label{fig:eaxt_fig1}
	\end{center}
\end{figure}

\begin{figure}[t]
	\begin{center}
		\includegraphics[width=8.5cm]{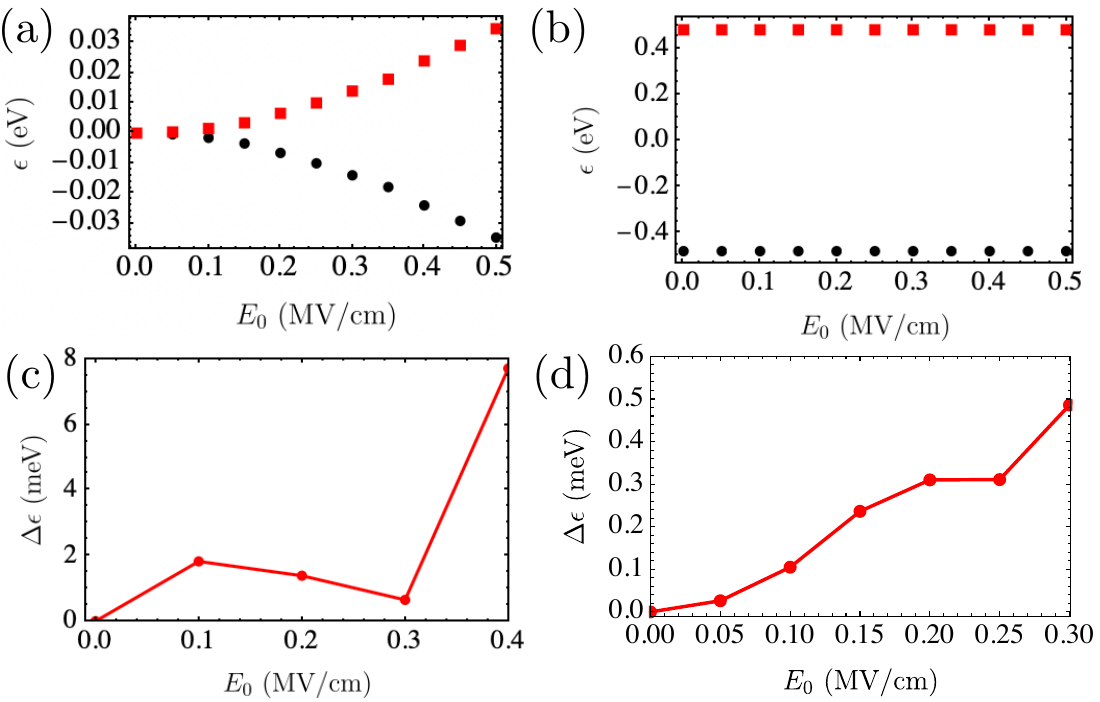}
		\caption{(Color online) Light-driven bilayer graphene quasienergy difference at the (a) $-K$ and (b) $\Gamma$ points as a function of the peak electric field for a pulse duration $\tau = 0.3$~ps. The quasienergy gap when considering both driven phonons and electrons is shown in panel (c), while (d) shows the gap when considering only driven phonons.}
		\label{fig:eaxt_fig2}
	\end{center}
\end{figure}

In Fig. \ref{fig:eaxt_fig1}, we plot the quasienergy spectrum in the vicinity of the $-K$ point for bilayer graphene driven with a laser pulse of duration $\tau = 0.3$~ps, $\Omega = 0.2$~ eV in resonance with the $E_u$ phonon mode, and a set of values of the peak electric field $E_0$. As $E_0$ increases, the quasienergy gap shows a non-monotonous behavior. To elucidate the behavior of this gap as a function of the peak electric field, in Fig. \ref{fig:eaxt_fig2}, we plot the quasienergies for the two central bands at the $-K$ and $\Gamma$ points in panels (a) and (b). At the $-K$ point, the applied laser pulse induces a splitting of the quasienergies as the peak electric field increases. On the other hand, at the $\Gamma$ point, presents a negligible effect ($|\epsilon(\Gamma)| \approx 0.4812$~ eV). The quasienergy gap near the Floquet zone center is shown in Fig. \ref{fig:eaxt_fig2} (c) as a function of the peak electric field, where we consider the effect of driven phonons and electrons. The behavior is non-monotonic. For comparison, in Fig. \ref{fig:eaxt_fig2} (d), we show the quasienergy gap obtained when considering only driven phonons. For the short pulse considered, the change in the quasienergy induced by the laser is dominated by the direct coupling with the electrons.

\subsection{Diabaticity of the electronic dynamics}

In the previous section, we discussed the exact solution of the time-dependent Schr\"odinger equation, $| \psi_\alpha (t) \rangle$, from a Floquet perspective. In this section, we compare the exact solution with the adiabatic Born-Oppenheimer (ABO) approximation, $\ket{\psi^{AB0}_{\alpha}(t)}$, defined as the solution of the eigenvalue equation
\begin{eqnarray}\label{eq:FH0}
\hat H(s)\ket{\psi^{AB0}_{\alpha}(s)} = E_{\alpha}(s)\ket{\psi^{AB0}_\alpha(s)},
\end{eqnarray}
where $E_{\alpha}(s)$ and $\ket{\psi_\alpha(s)}$ are the instantaneous energies and eigenstates for the rescaled time $s = \Omega t$. Since momentum remains a good quantum number in the presence of the  $\Gamma$-point phonon we consider in this work, we will consider as our initial state $\ket{\psi_{0}}=\ket{\psi^{AB0}_{\alpha={-K,2}}(s=0)}$. This corresponds to a state with momentum $-K$ and band index $n=2$. In the numerical calculations, we introduce an infinitesimal symmetry breaking term that allows to define unambiguously the band index. 
\begin{figure}[t]
	\begin{center}
		\includegraphics[width=8.5cm]{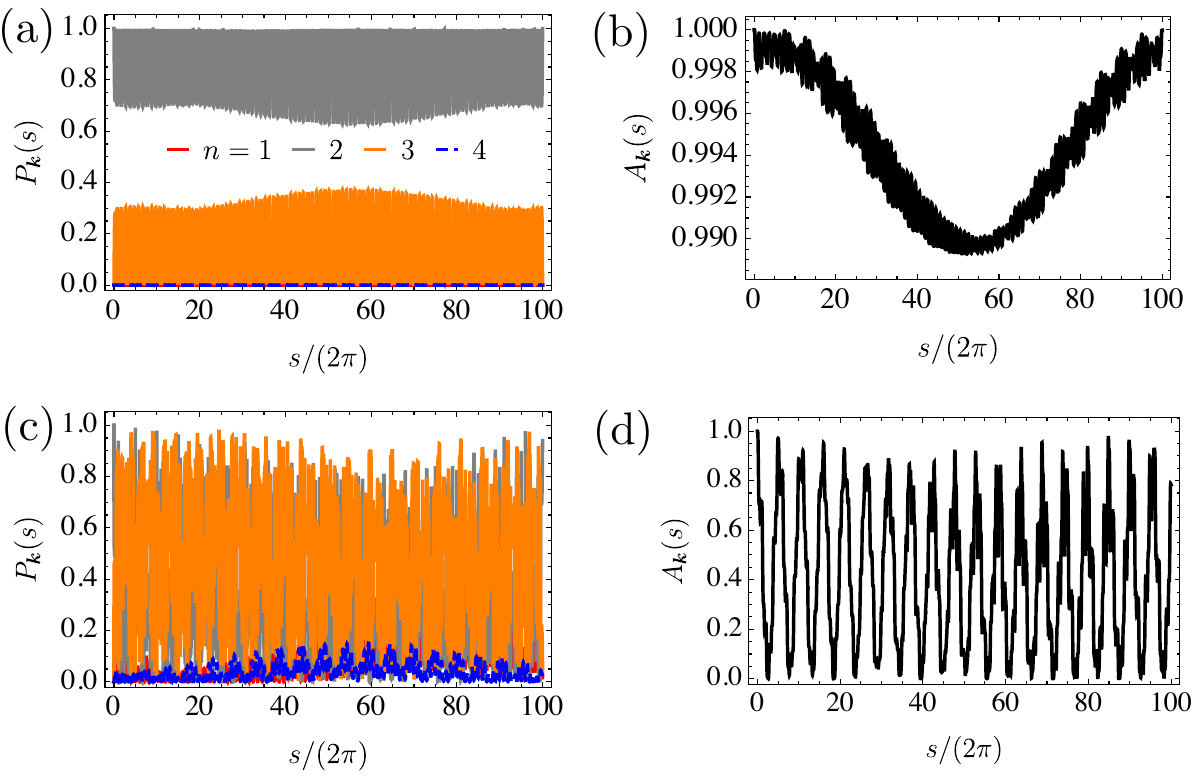}
		\caption{(Color online) (a) Overlap probability $P_{n}(s)$ of the ABO approximation with the exact electronic state following laser excitation of the phonons for $E_0=0.35$~MV/cm and $\tau = 0.8$ps at the $-K$ point. Gray and orange colors indicate bands $n=2$ and $n=3$ respectively. The overlap probability for bands $n=1$ and $n=4$ remains zero. (b) Autocorrelation function $A_k(s)$ for the exact electronic state. In panels (b) and (d), we show the corresponding probability $P_{n}(s)$ and autocorrelation for the case where electrons couple to the laser directly.}
		\label{fig:eaxt_fig3}
	\end{center}
\end{figure}
The deviation of the ABO approximation solution from the exact electronic wavefunction is quantified by the overlap probability $P_{n}(s) = | \langle \psi(s) | \psi^{AB0}_{n}(s) \rangle |^2$. We obtain the exact electronic wavefunction by solving the time-dependent Schr\"odinger equation exactly within a numerical approach. As an example, we consider the dynamics resulting from a laser pulse with peak electric field $E_0=0.35$~MV/cm and duration of $0.8$~ps. In Fig. \ref{fig:eaxt_fig3}(a), we show $P_{n}(s)$ assuming first that only phonons couple to the laser. In less than one full cycle ($s/(2\pi) = 1$) the probability $P_{n=2}(s)$ deviates from unity, and $P_{n=3}(s)$ acquires non-zero values, indicating the diabatic nature of the exact electronic wavefunction  $| \psi_ (t) \rangle$. This result is expect from the previous section's analysis, which revealed the formation of an electronic gap in the quasienergy spectrum. 

Figure \ref{fig:eaxt_fig3}(b) shows the square modulus of the autocorrelation $A_k(s) = |\langle \psi(s) | \psi_{0} \rangle|^2$, which quantifies the preservation of the electronic character, defined in Ref~\cite{Mohanty18316}. For the first five cycles ($s/(2\pi) = 5$), the autocorrelation remains close to unity, preserving the electronic state character for several vibrational cycles. This behavior suggests a diabatic time evolution of the electronic states following laser excitation, when only the lattice vibrations are considered. This result is expected following the study presented in Ref. \cite{Mohanty18316}, where thermally-excited phonons leads to a diabatic time-evolution and preservation of the electronic initial state character. Panels (c) and (d) show $P_{n}(s)$ and $A_k(s)$ taking into account the direct coupling of the electron with the laser. While the ABO approximation also deviates from the exact electronic wavefunction, the exact electronic wavefunction does not retain its initial state character, as shown by the deviation of $A_k(s)$ from unity in one cycle. Thus, although both processes (driven electrons and driven phonons) occur at the same frequency, their impact on the electronic dynamics is completely different. 

\begin{figure}
	\begin{center}
		\includegraphics[width=8.5cm]{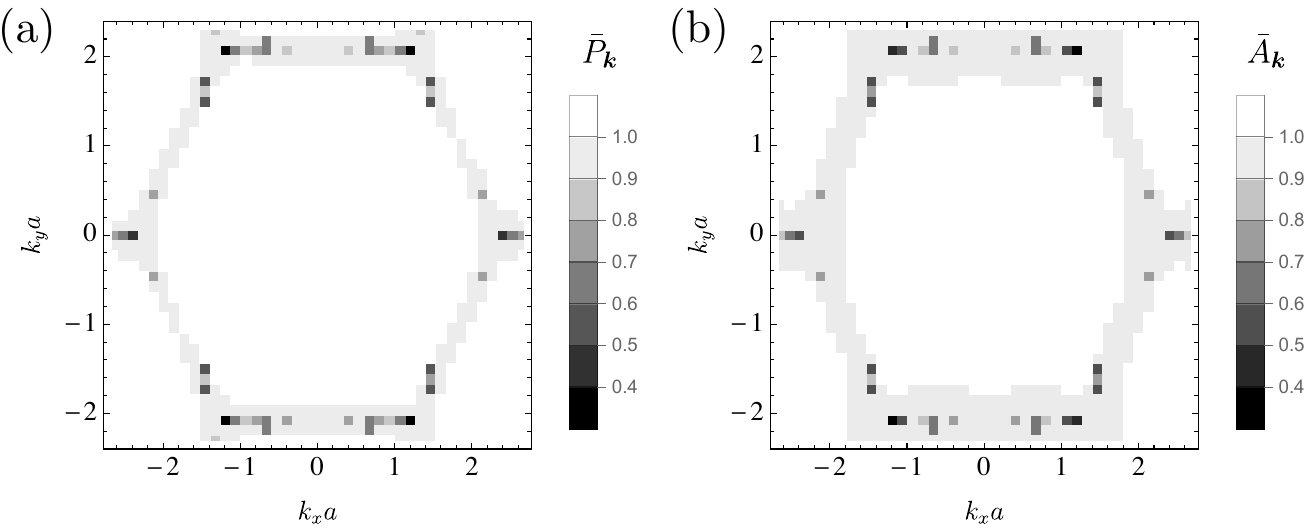}
		\caption{(Color online) (a) Averaged overlap probability $\bar P_{n=2}$ map in the Brillouin zone of the ABO approximation with the exact electronic state following laser excitation for $E_0=0.35$~MV/cm and $\tau = 0.8$ps. (b) Averaged autocorrelation function $\bar A_k$ for the exact electronic state.}
		\label{fig:eaxt_fig4}
	\end{center}
\end{figure}

At the $\Gamma$ point, the initial state $\ket{\psi_{0}}=\ket{\psi^{AB0}_{\alpha={\Gamma,2}}(s=0)}$ leads to probabilities $P_{n}(s) \approx \delta_{n,2}$, which supports an adiabatic time-evolution for electronic states away from the degeneracy points for both direct electronic and phononic coupling with the laser. In Fig. \ref{fig:eaxt_fig4} we show the diabaticity maps across the whole Brillouin zone. Deviations are mainly obtained only at the zone boundary due to the relative energy difference between the bilayer graphene bands in equilibrium.

\section{Conclusions}
\label{sec:conclusions}

In this work, we derived a time-dependent tight-binding model Hamiltonian for phonon- and electron-driven honeycomb bilayers with point group $D_{3d}$ assuming an ``atomically adiabatic" approximation. We applied our model to AB stacked bilayer graphene driven with infrared light, and solved the time-dependent model exactly, without invoking the adiabatic Born-Oppenheimer approximation. We showed that the exact electronic quasienergy spectrum develops a gap near the $-K$ point, in contrast with the expectation from the adiabatic approximation. We studied the behavior of the gap as a function of the peak electric field in the laser pulse, and compared the effect of driving the electrons and driving the phonons. Within our time-dependent tight-binding model, we find that the electronic gap is dominated by the effect of the direct coupling of the laser with the electrons and not by the vibrating nuclei. We analyzed the diabatic nature and autocorrelation of the electronic dynamics for both driven electron and driven phonons separately. In both cases, the adiabatic solution deviates from the exact solution within one cycle. However, the electronic autocorrelation showed contrasting behavior. When driving only the phonons, the exact solution remains highly correlated with the initial state. This result is in agreement with previous studies on thermal phonons in graphene. When including the effect of the laser on the electrons, the exact electronic wavefunction quickly deviates from the initial state. The procedure here detailed to derive time-dependent tight-binding models for driven systems can be applied to other van der Waals materials to uncover the interplay of driven phonons and electrons in the low-frequency regime, accessible in current experimental setups.

\section{Acknowledgements}

We thank Mikel Iraola, Maia G. Vergniory, and Stepan S. Tsirkin for useful discussions about the IrRep code. This research was primarily supported by the National Science Foundation through the Center for Dynamics and Control of Materials: an NSF MRSEC under Cooperative Agreement No. DMR-1720595. We acknowledge additional support under NSF DMR-1949701 and DMR-2114825. M. R-V. was supported by LANL LDRD Program and
by the U.S. Department of Energy, Office of Science, Basic Energy Sciences, Materials Sciences and Engineering Division, Condensed Matter Theory Program. The authors acknowledge the Texas Advanced Computing Center (TACC) at The University of Texas at Austin for providing HPC resources that have contributed to the research results reported within this paper. URL: http://www.tacc.utexas.edu.

\appendix 

\section{Electronic band representations}
\label{app:br}

Bilayer graphene corresponds to the space group $P\bar{3}m1$ (No. 164). The symmetry operations in Seitz notation and in the Standard setting are~\cite{BCS}: 

\begin{itemize}
    \item[1)] $\left\{1 \mid 0\right\} = \left(\begin{array}{llll}1 & 0 & 0 & 0 \\ 0 & 1 & 0 & 0 \\ 0 & 0 & 1 & 0\end{array}\right)$
    \item[2)]  $\left\{-1 \mid 0\right\}=
    \left(\begin{array}{llll}-1 & 0 & 0 & 0 \\ 0 & -1 & 0 & 0 \\ 0 & 0 & -1 & 0\end{array}\right)$
    \item[3)] $\left\{3_{001}^{+} \mid 0\right\} = \left(\begin{array}{rrrr}0 & -1 & 0 & 0 \\ 1 & -1 & 0 & 0 \\ 0 & 0 & 1 & 0\end{array}\right)$
    \item[4)] $\left\{-3_{001}^{+} \mid 0\right\} = \left(\begin{array}{rrrr}0 & 1 & 0 & 0 \\ -1 & 1 & 0 & 0 \\ 0 & 0 & -1 & 0\end{array}\right)$
    \item[5)] $\left\{3_{001}^{-} \mid 0\right\} = \left(\begin{array}{rrrr}-1 & 1 & 0 & 0 \\ -1 & 0 & 0 & 0 \\ 0 & 0 & 1 & 0\end{array}\right)$
    \item[6)] $\left\{-3_{001}^{-} \mid 0\right\} = \left(\begin{array}{rrrr}1 & -1 & 0 & 0 \\ 1 & 0 & 0 & 0 \\ 0 & 0 & -1 & 0\end{array}\right)$
    \item[7)] $\left\{2_{110} \mid 0\right\} = \left(\begin{array}{llll}0 & 1 & 0 & 0 \\ 1 & 0 & 0 & 0 \\ 0 & 0 & -1 & 0\end{array}\right)$
    \item[8)] $\left\{m_{110} \mid 0\right\} = \left(\begin{array}{llll}0 & -1 & 0 & 0 \\ -1 & 0 & 0 & 0 \\ 0 & 0 & 1 & 0\end{array}\right)$
    \item[9)] $\left\{2_{100} \mid 0\right\}= \left(\begin{array}{rrrr}1 & -1 & 0 & 0 \\ 0 & -1 & 0 & 0 \\ 0 & 0 & -1 & 0\end{array}\right)$
    \item[10)] $\left\{m_{100} \mid 0\right\}= \left(\begin{array}{rrrr}-1 & 1 & 0 & 0 \\ 0 & 1 & 0 & 0 \\ 0 & 0 & 1 & 0\end{array}\right)$
    \item[11)] $\left\{2_{010} \mid 0\right\}= \left(\begin{array}{rrrr}-1 & 0 & 0 & 0 \\ -1 & 1 & 0 & 0 \\ 0 & 0 & -1 & 0\end{array}\right)$
    \item[12)] $\left\{m_{010} \mid 0\right\}= \left(\begin{array}{rrrr}1 & 0 & 0 & 0 \\ 1 & -1 & 0 & 0 \\ 0 & 0 & 1 & 0\end{array}\right)$
\end{itemize}

The first three columns (from left to right) define the three-dimensional rotation matrix. The last column corresponds to translations. The $\Gamma$ point posses the 12 symmetries listed above. The $K$ point has symmetries $ 1 ,       3,        5,        7,        9,       11  $. Finally, the M point posses symmetries $ 1,        2,       11,       12$.

\section{Bilayer graphene band structure from first principles}
\label{sec:bands}
To determine the electron and phonon properties, we employ density functional theory (DFT)~\cite{kohn1965} as implemented in {\sc QUANTUM ESPRESSO}~\cite{QE-2017, QE-2009, doi:10.1063/5.0005082}. We two approximation schemes for the exchange and correlation potentials: 1) the local density approximation (LDA)~\cite{perdew1981}; and 2) the generalized-gradient approximation (GGA) ~\cite{perdew1996} supplemented with van der Waals interactions~\cite{doi:10.1002/jcc.21112,doi:10.1002/jcc.20495}. In both cases, we consider projector augmented-wave (PAW) pseudo-potentials~\cite{kresse1999,blochl1994}. We introduce a vacuum of $15 \AA$ to avoid interaction effects between supercells. 

First, we perform a variable cell relaxation calculation to obtain our optimal lattice parameters. We employ a kinetic energy cutoff for the wave functions of 45 Ry and 540 Ry for the charge density. The Brillouin zone (BZ) was sampled with a $34 \times 34  \times 1 $ grid, dense enough for the convergence of our results. We use a smearing for the Fermi distribution of $10^{-3}$ Ry~\cite{marzari1999}. The forces between the atoms were converged to $1$ meV/\AA, and the total energy to $0.05$~meV. Table \ref{tab:a_and_d} shows our result for the lattice constants and separation between the graphene layers. The lattice constant $a$ and distance between the layers $d$ are in good agreement with other equivalent DFT calculations~\cite{min2007,ulman2014} and experiments~\cite{doi:10.1098/rspa.1924.0101}. In Fig. \ref{fig:bands_gga}, we show the band structure along a high-symmetry path in the BZ. Near the $K$ and $K'$ points, we obtain quadratic band touching points at the Fermi level. 

\begin{table}[]
\centering
\begin{tabular}{|c|c|c|c|c|}
\hline
            & LDA & GGA & LDA Ref. \cite{ulman2014} & GGA Ref.\cite{ulman2014}  \\ \hline
$a$ (\AA)     &   2.44  &    2.46  &   2.44    & 2.46\\ \hline
$d$ (\AA)     &  3.30   &    3.25  & 3.34      & 3.20 \\ \hline
\end{tabular}
\caption{Lattice constant and distance between the layers for bilayer graphene obtained with LDA and GGA approximations. Experimentally, $d=3.35$ (\AA) for graphite~\cite{doi:10.1098/rspa.1924.0101}. }
\label{tab:a_and_d}
\end{table}

\section{The problem with a time-dependent Wannier basis in the high frequency regime}
\label{app:dimer}
In this section we will consider a simplified problem that will allow us to understand what nonphysical behaviour can arise in the Floquet case if we consider a time-dependent Wannier basis and do not correct by a term $i(\partial_t P)P$.

\begin{figure}
	\centering
	\includegraphics[width=1\linewidth]{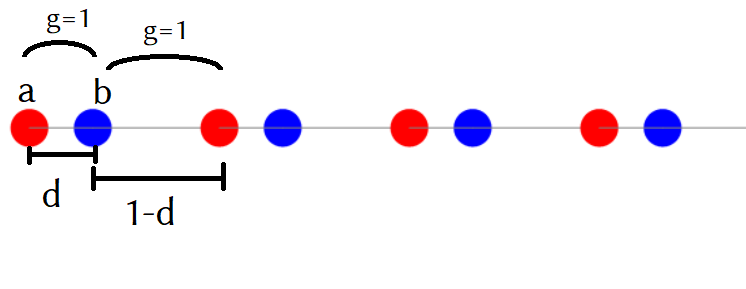}
	\caption{Dimer with period 1 and hopping strength $g=1$. The distance between atoms a and b in a unit cell is alternating as $d$ and $1-d$.}
	\label{fig:dimer}
\end{figure}

Let us consider the dimer depicted in figure \ref{fig:dimer}. 
We assume that electrons moving on this lattice are described by the simple tight binding Hamiltonian
\begin{equation}
	H=\sum_i a_i^\dag b_{i+d}+a_{i+1}^\dag b_{i+d}+h.c.
\end{equation}

In momentum space it can be rewritten as
\begin{equation}
	H=(a_k^\dag, b_k^\dag) h_k\begin{pmatrix}
	a_k\\b_k
	\end{pmatrix}
\end{equation}

with 
\begin{equation}
	h_k=\begin{pmatrix}
	0&e^{ikd}+e^{ik(d-1)}\\
	e^{-ikd}+e^{-ik(d-1)}&0
	\end{pmatrix}.
\end{equation}
 Now the eigenvalues one finds for $h_k$ are
 \begin{equation}
 	E=\pm 2\cos(k/2)
 \end{equation}
 and are periodic with period $4\pi$ of the Brillouin zone as one would expect regardless of value $d$.

Next we consider the case where $d=\cos(\omega t)d_0$ and the position of site b changes with time. In the adiabatic limit where $i(\partial_t P)P$ can be neglected we find that the instantaneous energies all have the periodicity of the Brillouin zone and therefore at first everything is fine. However, when we take the high frequency regime $\omega\to \infty$ where $i(\partial_t P)P$ cannot be neglected we will see totally different behaviour. 

In this limit the Floquet Hamiltonian is just given by the Hamiltonian time averaged over one period and one finds that
\begin{equation}
	h^{F,\omega\to\infty}_k=\begin{pmatrix}
	0&\left(1+e^{-i k}\right) J_0(d_0 k)\\
	\left(1+e^{i k}\right) J_0(d_0 k)&0
	\end{pmatrix}
\end{equation}

For the eigenvalues one obtains
\begin{equation}
	E=\pm 2 \cos \left(\frac{k}{2}\right) J_0(d_0 k).
\end{equation}
Therefore, one finds the nonphysical result that the quasienergies are not periodic in momentum space anymore despite the system being periodic at all times. 

This effect is also present in the tight-binding model studied in the main text. In Fig. \ref{fig:bz}, we plot the a quasienergy band in the first Floquet zone for more than one Brillouin zones to reflect the momentum space periodicity for a set of representative parameters. In the high-frequency regime, the periodicity is lost, while in the low-frequency regime (not necessarily the adiabatic limit) the periodicity is recovered.

\section{Low-frequency momentum space periodicity for driven bilayer graphene}
\label{app:bz-recovery}

\begin{figure}
	\centering
	\includegraphics[width=1\linewidth]{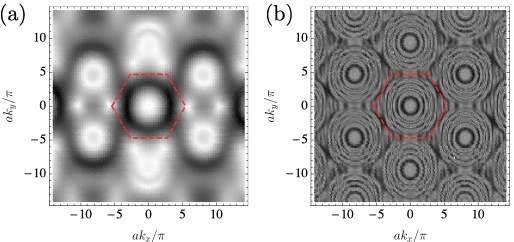}
	\caption{(a) Momentum space periodicity is lost in the high-frequency regime for driven bilayer graphene. In the low-frequency regime (b), the periodicity is recovered. The red dashed line corresponds to the first Brillouin zone.}
	\label{fig:bz}
\end{figure}

In the main text, we showed that the full time-dependet tight-binding Hamiltonian is given by $H_{\mathrm{TB}}=i(\partial_t P)P+PHP$, where $P$ is the projection operator $P$ onto the suitable Wannier orbitals~\cite{PhysRev.133.A171}. One of the most striking effects of the neglecting the term $i(\partial_t P)P$ appears in the loss of momentum space periodicity in the high-frequency regime. In Fig. \ref{fig:bz} we show the Fourier transform in the (a) high-frequency regime ($\hbar \Omega \approx 10$~eV), and in the (b) low-frequency regime ($\hbar \Omega \approx 1$~eV). The phonon drive parameters is $\Delta /a = 0.8$, much stronger than a realistic lattice distortion achievable in experiments, $\Delta /a = 0.05$. The laser coupling with the electrons is not considered since it does not introduce time-dependence in the projector operator $P$. 

%

\end{document}